\documentclass{article}
\usepackage{spconf,amsmath,graphicx,multirow,enumitem}

\title{Supervised and Unsupervised Approaches for Controlling Narrow Lexical Focus in Sequence-to-Sequence Speech Synthesis}
\name{Slava Shechtman$^1$, Raul Fernandez$^2$, David Haws$^2$}
\address{
  $^1$IBM Haifa Research Lab, Haifa -- Israel \\
  $^2$IBM TJ Watson Research Lab, Yorktown Heights, NY -- USA\\
{\small \tt slava@il.ibm.com, \{fernanra, dhaws\}@us.ibm.com}}

\begin{document}
\ninept
\maketitle
\begin{abstract}
% 100-150 words
Although Sequence-to-Sequence (S2S) architectures have become state-of-the-art in 
speech synthesis, capable of generating outputs that
approach the perceptual quality of natural samples, they are limited by
a lack of flexibility when it comes to controlling the output.
In this work we present a framework  capable of controlling the prosodic
output via a set of concise, interpretable, disentangled parameters. We apply
this framework to the realization of emphatic lexical focus, proposing a variety of
architectures designed to exploit different levels of supervision based on
the availability of labeled resources. We evaluate these approaches via listening
tests that demonstrate we are able to successfully realize controllable focus
while maintaining the same, or higher, naturalness over an established
baseline, and we explore how the different approaches compare
when synthesizing in a target voice with or without labeled data. 
\end{abstract}

\section{Introduction}
\label{sec:intro}

Sequence-to-Sequence (S2S) speech-synthesis architectures have become the state-of-the-art in the field, 
providing high-quality outputs that approach or match
the perceived quality of natural speech in many studies. Aside from the level of quality attained, there are many attractive 
features to these models. They are able to jointly model different aspects of a waveform (e.g., segmental and prosodic), so interactions between
them can be implicitly learned. They also do away with classical pipeline architectures in favor of a single unified model, which is
appealing when some of the modules in the pipeline are difficult to develop (e.g., text-processing for a new language).
On the other hand, they suffer from well-documented shortcomings, such as lack of interpretability (it can be difficult to tell which parts of the model are responsible 
for what functions), lack of controllability (it is more difficult to intervene into the model in order to control some aspects of the synthesis, which is often desired,
such as when providing SSML support), and potential instability (small deviations at inference time can become exacerbated and generate highly 
degraded speech). 

In this work, we address the {\em controllability} issue by expanding the S2S architecture with mechanisms that can be exposed to the user 
to manipulate some property of the output. Although usability factors are not the focus of this work, 
we nonetheless advocate for a set of properties that will 
make such controls accessible to the end consumer of the system, namely:
\begin{itemize}[leftmargin=*] 
\item{Interpretability}: The listener should be able to clearly hear and identify the effect of varying a control (e.g., speech is slower, faster, higher-pitched, sounds happier, etc.). 
\item{Monotonicity}: A design that results in perceptual effects that vary monotonically as the user varies a control has a more
intuitive feel, and is more easily tunable.
\item{Low-dimensionality}: The user should not be expected to manipulate a large number of parameters to control the output. 
The model should  either expose a low-dimensional controllable representation, or be able to step in and fill in defaults to obviate the task 
for the user.
\item{Disentanglement}: Though this may be difficult due to the many 
ways different speech parameters interact, a set of controls that are more decoupled
from each other facilitates the tuning of the output along fairly independent (perceptual) 
dimensions (e.g., tempo and volume could be tuned separately without needing to revisit a previously
tuned parameter).
\end{itemize}
We explore the realization and controllability of narrow lexical focus as a case study for the above. Our objective is the
realization of an emphatic level of prominence that is distinct from the type of accentuation that we observe
in ``neutral'' broad-focus prosody. Consider the intonational phrase in the examples below when they
occur as a reaction to the context in parentheses. In E1, as a reply to a general question, we see a likely case of broad focus
prosody, where {\em wine} acts as the nuclear element and receives some sort of pitch accent. The same
accented word, however, might be given a more emphatic degree of prominence when it happens in the context of E2. Furthermore,
we can switch the focal point to a different word in the phrase when it is primed by a different context, as in E3. The {\em [...]} in these
examples delimit the domain of focus, which the speaker may delineate, for instance, by employing a higher degree of disjuncture
between the focal element and its context.
\begin{itemize}[leftmargin=*] 
\item{E1:} {\em Mary is [pouring the wine].} ({\em What's Mary doing?})
\item{E2:} {\em Mary is pouring the [\underline{wine}].} ({\em Is Mary pouring the beer?})
\item{E3:} {\em [\underline{Mary}] is pouring the wine.} ({\em Is John pouring the wine?})
\end{itemize}
We are interested in the prosodic realizations that arise in examples such as E2 and E3 above (but also in other 
scenarios, such as contrastive emphasis, requesting clarification, etc.).
In Sec.~\ref{sec:model} we introduce an S2S architecture
that supports this type of prosodic control, review in Sec.~\ref{sec:previous} how our approach compares 
to relevant research in the literature, evaluate competing approaches to this question in Sec.~\ref{sec:eval},
and conclude in Sec.~\ref{sec:discuss} with some analysis of these results and an outline of future steps.

% TODO: Find a way to address this from R3 (in S1 or S2?):
% It involves too much guessing in section 1 before reading section 2.  It's easier for the readers to know the definition of these two terms:
%- indicator feature (label?)
%

\section{Architecture}
\label{sec:model}

\begin{figure}[hbt]
  \centering
\includegraphics[width=\linewidth]{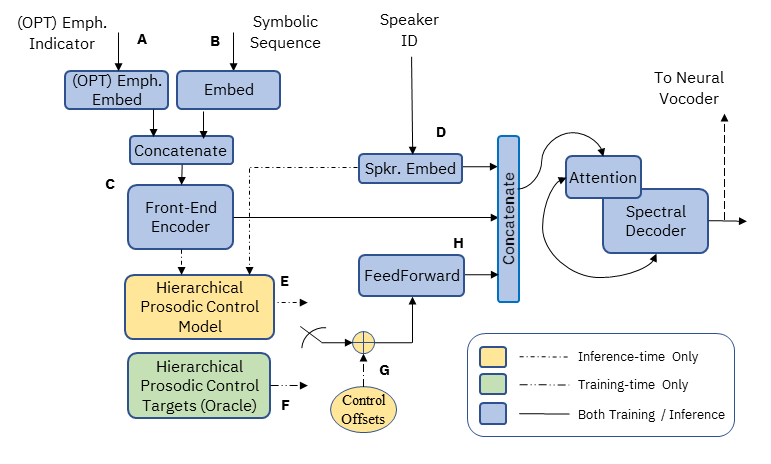}
  \caption{System architecture. The dashed line indicates the output of the S2S model is sent to a separately trained neural vocoder which does not play a role in the 
optimization of Eqn.~\ref{eq:loss}.}
  \label{fig:arch}
\end{figure}

The model (Fig.~\ref{fig:arch}) is a variant of the Tacotron2 architecture  proposed in~\cite{Shen-Pang:18}, 
augmented with components in the  decoder
to facilitate both the injection of controls, and improved stability during decoding~\cite{Shechtman-Rabinovitz:20}. 
This sequence-to-sequence model generates an acoustic spectral-prosodic representation that is then fed to an independently-trained, 
LPC-Net-based~\cite{Valin-Skoglund:19} 
neural vocoder to generate high-quality samples in real time~\cite{Shechtman-Rabinovitz:20}.

The {\bf Encoder} comprises the following components, combined as in Fig.~\ref{fig:arch} before being sent to the decoder:
\begin{itemize}
\item  The {\em emphasis embedding} (A) from a Boolean indicator feature encoding emphatic focus within the utterance, as a way to 
provide direct supervision to the model.
\item The {\em embedding of various linguistic symbols} (B) extracted from an extended phonetic dictionary comprising
phone identity, lexical stress, phrase type, and other symbols for word boundaries and silences. This analysis is carried out externally 
by a rules-based TTS Front End module,  adopted from a unit selection system~\cite{Pitrelli:06}.  
\item A {\em front-end encoder} (C) consisting of convolutional and bi-directional Long Short-Term Memory (Bi-LSTM) layers (as in~~\cite{Shen-Pang:18}), 
encoding the merged embeddings from (A) and (B). 
\item A {\em global utterance-level speaker embedding} (D), broadcast over the length of the sequence, to support training in a multi-speaker setting.
\item A set of 4-dimensional {\em hierarchical prosodic controls} (which will be introduced in Sec.~\ref{ssec:proscontrol}) designed to
enable the type of fine, word-level  modification needed to realize the prosodic patterns associated with emphatic focus.
Since these prosodic controls are a set of statistics extracted from the acoustic signal, the ground-truth values from the training
set are used during  training (F). At inference time a separate predictive module (E) steps in to provide default predictions for the hierarchical prosodic trajectories.
\item An optional {\em user-exposed control} (G) to modify the default predictions generated by (E).
In particular, we propose a set of additive controls that are linguistically intuitive and interpretable (Sec.~\ref{ssec:proscontrol}). 
Note that the feed-forward operation in block H is placed {\em after} the (optional) user request in G. This
design choice is made to preserve the interpretability of the quantities the user gets to manipulate (which would not be the case if the order
was reversed, and the independent prosodic targets
were blended via a non-linear feed-forward operation).
\end{itemize}

The {\bf Decoder}  is an autoregressive network that largely follows the standard Tacotron2 architecture, 
but with modifications on the attention mechanism, autoregressive feedback, choice of targets,
and training losses. These have already been described in~\cite{Shechtman-Rabinovitz:20}
%~\cite{Shechtman-Sorin:19} 
and are summarized 
here as follows. The attention is an {\em augmented two-stage attention} where the content- and location-based attention of
Tacotron2  are followed by a structure-preserving  mechanism encouraging monotonicity and unimodality in the alignment matrix. 
This modification has been found to be crucial to increase stability during inference, particularly in the presence of
external controls. A double feedback approach is used during training to expose the model both to the previous ground-truth output value (i.e., teacher forcing) as
well as the previous predicted value (i.e., inference mode). At inference time, the predicted value is replicated.
The model is {\em trained in a multi-task fashion} to predict the 80-dim mel cepstral features in tandem with the parameters needed as inputs for an
independently trained
LPC-Net neural vocoder. For 22kHz signals, these features (which we denote as ``LPC features'') consist of a
22-dim vector with 20 cepstral coefficients, log $f_0$ and $f_0$ correlation. The predicted LPC features are also processed with two post-nets 
(one to refine the cepstrum, and one to refine the pitch parameters); no post-net refinement is applied on the mel task. 
Let $y^M_t$ and $y^L_t$ represent the target sequences for the mel and LPC tasks respectively,
$\tilde{y}^M_t$ and $\tilde{y}^L_t$ their final predictions, and $\hat{y}^L_t$  the ``intermediate'' LPC-feature
prediction (before the post-net). Then the following differential loss function is used to train the system:
\begin{align}
\mathcal{L} 	&	= MSE(\tilde{y}^M_t, y^M_t) + 0.8MSE(\hat{y}^L_t, y^L_t)  \nonumber \\
			& + 0.4MSE(\tilde{y}^L_t, y^L_t) + 0.4MSE(\Delta\tilde{y}^L_t, \Delta y^L_t  ),
\label{eq:loss}
\end{align}
where the $\Delta$ operator applies the first difference in time to a sequence, and $MSE(,)$ is the mean-squared error.
For the sake of space, we omit some detail in this exposition, and refer the reader to~\cite{Shechtman-Sorin:19,Shechtman-Rabinovitz:20} 
for additional background and formulae.

This architecture accommodates some variants depending on the availability of labeled resources and types of control that are exposable to a  user. 
Among them,  we explore the following:
\begin{itemize}[leftmargin=*] 
\item {\em Classic Supervision}: When labeled data is available, this architecture conditions directly on a Boolean indicator feature. During training, the ground truth values of the audio signals are used. During
inference a binary request is passed to the system\footnote{This value could be either user-specified for given words (e.g., via mark-up) or inferred from text. 
We do not address here the problem of 
inference from text (though we have previously in~\cite{Mass-Shechtman:18}),
 and focus on the realization of prosodic controls assuming an existing request.}. This corresponds to blocks \{A, B, C, D\} on the encoder side.
\item {\em No Supervision:} Under the assumption that {\em no} labeled data exists, the architecture defined by \{B, C, D, E, F, G, H\}
provides a way to
introduce sensitivity into the S2S system during training, and, at inference time, control the realization of the prosodic patterns via a {\em tunable} 
set of controls (cf. the binary control of the supervised architecture).
\item {\em Hybrid:} Though components  E through H are motivated by an unsupervised approach, they may facilitate the realization of prosodic patterns
{\em even} when labeled data exists by working in tandem with an explicit feature. To investigate this, we consider
 a ``hybrid'' approach (defined by the full model \{A-G\})
that mixes supervised knowledge with the infrastructure designed to tackle the case when we don't have access to it.  
\end{itemize}

\subsection{Hierarchical Prosodic-Control Model}
\label{ssec:proscontrol}
Following the motivation for a perceptually-interpretable, low-dimensional control mechanism for prosody discussed in Sec.~\ref{sec:intro}, we propose a
hierarchical set of four prosodic controls that summarize information about the duration and pitch excursion of a signal over linguistically-meaningful and intuitive intervals of
the prosodic hierarchy. These controls include global and local properties, and are  an extension of the approach 
in~\cite{Shechtman-Sorin:19}, which allowed for controlling global aspects like overall tempo, but which lacked any control to effect the kind of
deviation from long-term trends needed to realize local emphatic focus. To arrive at these, let us first define the following statistics:
\begin{itemize}[leftmargin=*] 
\item $S_{dur}$: The log of the average per-phone durations, along a sentence (and excluding any silence).
\item $S_{f_0}$: The log-$f_0$ ``spread'' (defined as the difference between the 95- and 5-percentiles of log-$f_0$), along a sentence.
\item $W_{dur}$: The log of the average per-phone durations (as above), along each word.
\item $W_{f_0}$: The log-$f_0$ ``spread'' (as above), along each word.
\end{itemize}
Note that the average per-phone durations in the above definitions are estimated as the duration of speech (in seconds) along the relevant spans (word or sentence) divided by
the number of phone symbols contained therein, and that therefore no fine-level phonetic alignment is required in the computation (only coarse word-level alignments
and either phonetic transcriptions or a dictionary).
These sentence- and word-level properties are propagated down to the temporal granularity of the phonetic encoder outputs (i.e., phones) to form piecewise functions that are constant within a (sentence
or word) unit.
%(Fig~\ref{fig:prsinfo}). 
From this we define the following four-component prosodic-control target vector:
\begin{equation}
PC = Norm_\sigma\{ [S_{dur}, S_{f_0}, W_{dur}-S_{dur}, W_{f_0}-S_{f_0}]\},
\label{eq:prsinfo}
\end{equation}
where $Norm_\sigma\{\}$  is the linear map $[-3\sigma^2,3\sigma^2]\rightarrow [-1, 1]$, and $\sigma^2$ is the global (corpus-wide) variance for each of the
statistics in $PC$.
At inference time, the predictions of the prosodic-control subnet are rectified to be piecewise 
constant as the oracle  values that the S2S system was trained with.  In the evaluated
systems, a mean pooling function is applied to the prediction to be constant between the (known) sentence and word boundaries.

\begin{figure}[htb]
  \centering
  \includegraphics[width=\linewidth]{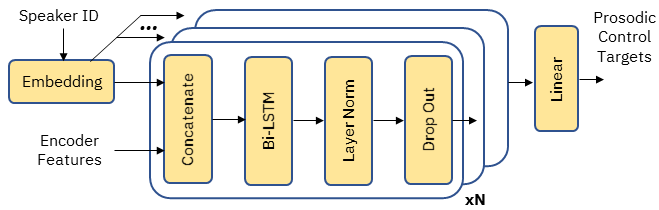}
  \caption{Architecture of the hierarchical prosodic sub-network for predicting targets from encoder-level features.}
  \label{fig:subnet}
\end{figure}

The architecture of the prosodic-control predictor (Fig.~\ref{fig:subnet}) consists of a stack of N blocks, each comprising a
concatenation of the speaker embedding with the block's input,  a Bi-LSTM, Layer normalization~\cite{Ba-Kiros:16}, and Drop-Out.
Models are trained in a multi-speaker fashion via a speaker-embedding layer whose output is fed into every cascaded block.
(We will discuss how we instantiate model sizes for the different components of this architecture when we discuss the details of
selecting models for evaluation in Sec.~\ref{sec:eval}.) Since the replication to the phone level artificially introduces an over-contribution to the loss, each observation in each of the 
prosodic targets
is down-weighted by this replication factor (e.g., for the sentence-level targets, each phone-level observation in a 10-phone sentence receives a weight of $0.1$; 
a similar approach is applied to the word-level
targets). These observation-level weights (uniquely determined by prosodic constituency) are then combined with global target-specific 
weights $\alpha$ that can be set during training to
trade-off between the different targets (in this evaluation $\alpha=[1, 1, 1.5, 3.5]$). 
The model is then trained with ADAM~\cite{Kingma-Ba:15} to minimize
the weighted L1 loss between predictions and targets. 
A set of 10\% of the sentences in the training set are held out to tune
structure (e.g., number of hidden units and blocks) and learning rate hyper-parameters .

\begin{figure}[htb]
  \centering
  \includegraphics[width=2.5in]{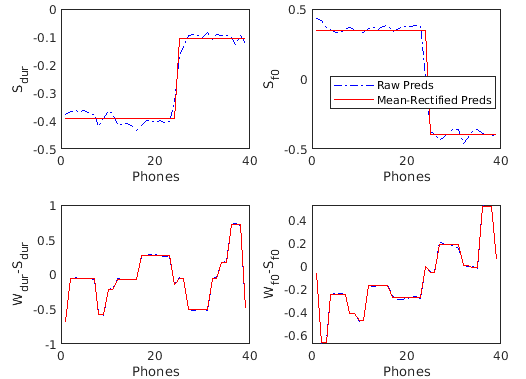}
  \caption{Sample phone-level trajectories of the four prosodic controls for a two-sentence input.}
  \label{fig:prsinfo}
\end{figure}

\begin{figure}[htb]
  \centering
  \includegraphics[width=3.0in]{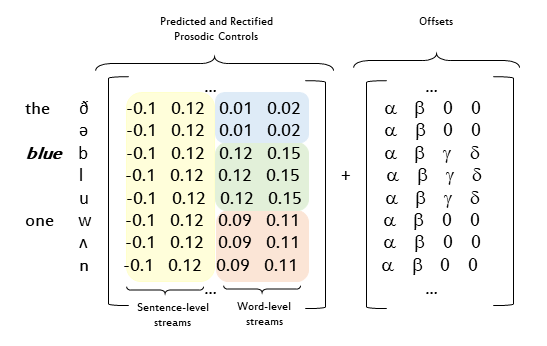}
  \caption{Boosting the prosodic-control predictions of sentence- and word-level targets to realize focus. The example shows a fragment of
an utterance where the word {\em blue} is to be emphasized (e.g., {\em I don't want the red one; I want the {\bf blue} one}). The
predicted prosodic controls are offset by global and local offsets, where the local offsets are applied to the focal words only.}
  \label{fig:offsets}
\end{figure}

At run time, lexical focus is controlled by the process illustrated in Fig.~\ref{fig:offsets}: The prosodic-control
predictions generated by component E in Fig.~\ref{fig:arch}, and post-processed to be piecewise constant, are offset by 4 tunable parameters 
$(\alpha,\beta,\gamma,\delta)$ where the $(\alpha,\beta)$ are global sentence-level offsets (that are applied uniformly and therefore only
contribute to the overall expressiveness of the utterance) and $(\gamma,\delta)$ 
boost the word-level predictions of {\em only} those words we wish to make salient (remaining non-focal words receive no offset).  These
run-time hyperparameters can be tuned via an independent development set.

\section{Previous and Related Work}
\label{sec:previous}

Synthesizing emphasis has been previously explored within other architectures like
unit selection~\cite{Raux-Black:03,Fernandez-Ramabhadran:07,Strom-Nenkova:07}, classical parametric synthesis~\cite{Yu-Mairesse:10,Meng-Wu:12,Do-Toda:16}, 
and pipeline systems using neural networks~\cite{Shechtman-Mordechay:18}. Within S2S models, controllability has recently received a moderate amount of attention,
with the Global Style Tokens (GST) proposal of~\cite{Wang-Stanton:18} being one of the earliest works to discover latent styles in an unsupervised fashion.
GST-based approaches have found wide usage (see, e.g.,~\cite{SkerryRyan-Battenberg:18,Lee-Kim:19,Valle-Li:20}), 
but as these representations are discovered, rather than explicitly formulated, they often lack {\em a priori} interpretability 
(though {\em post hoc} listening often reveals some uniform perceptual quality). 
GST and others~\cite{Park-Han:19} where global tempo is controllable
lack the finer-grained level of control we pursue due to its global nature. Non-GST approaches include works like~\cite{Zhu-Yang:19},
where direct conditioning on estimated indicators of emotion are used to control the output.
Recent work by~\cite{Ren-Ruan:19,Ren-Hu:20} has also looked at the controllability of prosodic properties in Transformer-based neural TTS systems, 
although at the core of  that approach is a move away from S2S models that, for the sake of speed, replaces a S2S teacher with feed-forward student 
models that decouple prosodic from spectral modeling.
We, in contrast, retain the full S2S framework in our implementation.
Hierarchical representations 
and controllability have been explored together in~\cite{An-Wang:19,Hsu-Zhang:19} though these 
approaches lack the level of interpretability for fine word-control.
The work of~\cite{Sun-Zhang:20} targets interpretable and controllable hierarchical prosodic controls and 
comes closest to the approach we pursue. However, their disentanglement is data-driven and leaves some residual couplings between
the (pitch and duration) dimensions we control separately; we model $f_0$ dynamics (as opposed to levels), which is
more perceptually relevant to realizing emphatic focus; and as we will see in the next section, our controllable systems attain
the same or higher level of quality when introducing this prosodic variation.
Prosody transfer across databases bearing different labels  is one of the main
applications of our framework, as we will discuss in Sec.~\ref{sec:eval}. The works of~\cite{Klimkov-Ronanki:19,Zhang-Pan:19}
pursue similar goals although, being based on global-sentence level embeddings,
they do not address fine-level control as we do.

\section{Evaluation}
\label{sec:eval}

The training material comprised four corpora from three professional native speakers of US English, broken down as follows: a set of~10.8K sentences from a male
speaker ($M1$); a set of~1K sentences from the same male speaker, where each sentence contains several emphasis-bearing words ($M1_{emp}$); and two
corpora from two distinct female speakers ($F1$ and $F2$) containing approximately~17.3K and~11K sentences respectively.
The corpus $M1_{emp}$ was collected by indicating to the speaker the emphasis-bearing words within each sentence, and instructing him to realize an emphatic level of prominence
on those target words. His prosodic realizations differ in marked ways from the style of broad focus prosody in terms of tempo, relative pitch accent height, and 
disjuncture from adjacent material.
The sentences were intended to serve as elicitors for various cases of narrow focus (e.g., contrast, disambiguation, etc.). 
Notice that labeled data is available for only one speaker, and that the size of this corpus is considerably smaller than that of the base corpora. 
A sentence from $M1_{emp}$ contains three emphatic words on average, and the overall percentage of such words was approximately~23\%. 
We define the following data partitions to facilitate the ensuing discussion: 
a set of data with all the resources pooled, including the emphatic data 
$D_{emp}=\{M1, 10\times M1_{emp}, F1, F2\}$, and a base set  $D_{base}  = \{M1, F1, F2\}$. Note that $D_{emp}$ uses 
10-fold replication of the $M1_{emp}$ subset to compensate for the lower prior.

We would like to investigate the trade-offs between approaches that use labeled data (when available), and the fully unsupervised approach that is
possible within the framework proposed in Sec.~\ref{sec:model}. To that end, consider the following systems:
\begin{itemize}[leftmargin=*] 
\item {\bf Base (NoEmph):} A baseline S2S system, which uses global (sentence-level) prosodic controls, but no word-level prosodic control.
The training set ($D_{emp}$) subsumes the emphatic data, but no other emphasis-marking feature is used.
\item {\bf Base (Sup):} A baseline system with {\em Classic Supervision} (as in Sec.~\ref{sec:model}) with global controls, 
trained with $D_{emp}$ and an explicit binary feature encoding the location of emphasis.
\item {\bf PC-Unsup:} A {\em Fully Unsupervised} system (as per Sec.~\ref{sec:model}) with variable prosodic control, 
where both the S2S and prosody-prediction components are trained 
with $D_{base}$.
\item {\bf PC-Hybrid:} A {\em Hybrid} system with variable prosodic control, trained with $D_{emp}$, 
and an explicit Boolean emphasis indicator as in the {\em Baseline (Sup)} model.
\end{itemize}

\begin{table}[thb]
\caption{Summary of the different properties and training strategies among the different systems evaluated.}
\label{table:systems}
\centering
\begin{tabular}{|c|c|c|c|c|} \hline
				& 	Base        & Base  		&  PC- 		& PC-		\\
                         &    (NoEmph) & (Sup)             & Unsup      &  Hybrid             \\\hline
Control?	&	N 							&	Y						& Y 				& Y			\\ \hline
Type of			&	\multirow{2}{*}{None}		& \multirow{2}{*}{Binary} 	& \multirow{2}{*}{Tunable}	& Binary  /	\\ 
Control?			&								& 							&							& Tunable 	\\ \hline
Training?		&  $D_{emp}$					& $D_{emp}$				& $D_{base}$				& $D_{emp}$ \\ \hline

Emph. feat? &	\multirow{1}{*}{N} 			& \multirow{1}{*}{Y} 		& \multirow{1}{*}{N} 		& 	\multirow{1}{*}{Y}\\ \hline
\end{tabular}
\end{table}

The architecture of {\bf Base (NoEmph)} with global controls was already presented and
evaluated in~\cite{Shechtman-Sorin:19}. Since it lacks fine-grained
lexical prosodic control, we do not expect it to perform well on an emphasis-evaluation task. 
It is used here, however, to provide a strong anchor point with respect to overall quality
to ensure that the alternative proposals do not degrade with respect to the naturalness afforded by this approach. 
A common LPC-Net neural vocoder, also trained in a multi-speaker fashion using $D_{base}$,
was used for all experiments~\cite{Shechtman-Rabinovitz:20}.

Model selection and tuning was done as follows. First, for the prosodic sub-network,  10\% of the training data was held out 
to do a grid search over structures and learning rate by tracking the held-out loss. 
The models thus selected were, for the {\bf PC-Unsup} condition, a stack of 5 blocks with 175 hidden units in the Bi-LSTM layer, 
and, for the {\bf Hybrid} model, a stack of 4 blocks with 200 hidden units in the Bi-LSTM layer. The speaker embedding was of
dimension 20 in both cases. Once this was fixed, a development set of 20 sentences not used in training 
was used to {\em perceptually} tune remaining hyper-parameters of the different configurations, 
including the dimension of the emphasis-embedding space ($dim=8$ for the {\em Hybrid} model, and $16$ for the {\em Base (Sup)} system),
and the run-time additive word-level boosting parameters $(\gamma,\delta)$  (see the ``control offset''
 component G in Fig.~\ref{fig:arch}, and Fig.~\ref{fig:offsets}) 
for the {\bf PC-Unsup} and {\bf PC-Hybrid} word-level controls (set to $(0.25, 1.30)$ and $(0.0, 1.5)$ respectively). These word-level offsets
were applied only to the item in a sentence that was intended to be the focus carrier; the predictions of the prosodic-control model remain
unboosted for all other lexical items.
Sentence-level boosting was not found to provide any advantages over word-level boosting, and the parameters $(\alpha,\beta)$ (see Fig.~\ref{fig:offsets})
were therefore only used for the
two reference systems (Base (NoEmph) and Base (Sup)), and set to $(0.0, 0.5)$). The non-negative boosting values we employ match our theoretical expectations,
and what we empirically observe in the $M1_{emp}$ subset, that focused items receive more pronounced pitch accents, and slower speaking
rate/longer durations. We observed that the {\em Base (Sup)} system already realized these tempo differences quite well, and only boosted the
pitch excursions when tuning the {\em Hybrid} systems. In general, we find that after tuning a single set of boosting parameters works quite well
across a variety of sentences and voices\footnote{Samples and additional listening test details are available at 
%https://s3.us-south.objectstorage.softlayer.net/zk-wav-data/Webpages/TTS-Taco-WordEmph\_SLT21/indexa.html
%https://s3.us-south.objectstorage.softlayer.net/zk-wav-data/Webpages/TTS-Taco-WordEmph-SLT21/indexa.html}
% available at http://ibm.biz/BdqM5X.}.
http://ibm.biz/SLT2021.}.

\subsection{Subjective Listening Tests}

We wish to evaluate how the different multi-speaker approaches we have described fare in a perceptual listening task. In particular, we are interested
in examining two test-case scenarios.  In the first case, we operate under the assumption that
the target synthesis voice matches a speaker for whom we have existing training data (i.e., the {\em matched} condition). In the second, and more
interesting case, we assume that the target synthesis voice lacks any such labeled resources for training (though some exists for a separate speaker), 
and that therefore any use the system makes of supervised information is done indirectly by transferring knowledge from one speaker to another 
(we refer to this as the {\em transplant} condition). Notice that the distinction we have just introduced applies to the systems that are sensitive to supervision in some way
(i.e., {\bf Base (Sup)} and {\bf PC-Hybrid}); system {\bf PC-Unsup}, by construction, is not.

To evaluate the systems defined in the previous section, while addressing the {\em matched} and 
{\em transplant} cases respectively, we conducted two independent listening tests
where the target speakers were $M1$ (whose training data contains an emphatic subset) and $F1$ (whose training data does 
not)\footnote{In informal listening, we found $F1$ and
$F2$ to be of comparable quality, so only one voice was selected to keep the test manageable.}. No natural recordings were included
 (which could have provided a topline performance) since no common set of utterances with emphasis existed for both voices, and
we wanted to run parallel tests. Instead, we opted for an evaluation set of 43 unseen sentences,
with each containing a single focused word.

\begin{table}[thb]
\caption{MOS ($\sigma$) results for the matched  condition ($M1$). For {\em emphasis} all systems are statistically 
significantly different from each other. 
For {\em quality}, there are no statistically significant differences between the 
pairs \{Base (NoEmph), PC-Unsup\} and \{Base (Sup), PC-Hybrid\}; 
all other pairwise differences are significant. Significance is assessed at the $p=0.01$ level via one-tailed t-tests.}
\label{table:lt-jjw}
\centering
\begin{tabular}{|c||c||c|} \hline
\multirow{2}{*}{System} & \multicolumn{2}{c|}{Attribute}	\\ \cline{2-3}
					& Emph 			& Quality \\ \hline\hline
Base (NoEmph)	& 2.21 (1.3)		& 3.87 (0.8) 		\\ \hline
Base (Sup)		& 4.08 (1.0) 		& 4.10 (0.1)		\\ \hline
PC-Unsup			& 3.35 (1.2)		& 3.82 (0.9)		\\ \hline
PC-Hybrid			& 3.96 (1.0)		& 4.08 (0.8)		\\ \hline
\end{tabular}
\end{table}

The listening tests were designed to evaluate the systems in terms of two attributes on 5-point scales: 
(i) how well they realize narrow focus on a given word, 
and (ii) the overall quality of the  sentence. Listeners were recruited through a crowd-sourcing platform
and presented with one audio sample at a time, accompanied by a transcript of the text where the intended focus-carrying word had been capitalized.
To facilitate comprehension of the task we provided the listeners with the following set of instructions, and collected their responses
in the provided 5-point scales:

\noindent {\em The UPPERCASE word (excluding the word "I", if it exists) in the text above should sound emphasized in this sample. 
Assess the {\bf level of emphasis} you hear in the UPPERCASE word. It sounds: 1 (neutrally spoken), 2, 3 (somewhat emphasized), 4, 5 (definitely emphasized). Assuming the  UPPERCASE word is emphasized as requested, rate the {\bf overall quality and naturalness} of this audio sample:
1 (Bad), 2 (Poor), 3 (Fair), 4 (Good), 5 (Excellent). }

Each \{sentence, system\} combination received 25 independent rating tuples (one for each of the 2 attributes). 
The texts were designed to make the choice of focus semantically congruent with the context-providing sentence. 
Tables~\ref{table:lt-jjw}-\ref{table:lt-sab} summarize the restuls in terms of Mean Opinion Scores (MOS), standard deviation
($\sigma$), and pairwise statistical significance.

\begin{table}[hb]
\caption{MOS ($\sigma$) results for the transplant condition ($F1$). All pairwise differences are statistically significantly
different for {\em emphasis}. For {\em quality}, \{Base (Sup.), PC-Unsup\} are statistically equivalent; all other pairwise differences
are statistically significantly different. Significance is assessed at the $p=0.01$ level via one-tailed t-tests.}
\label{table:lt-sab}
\centering
\begin{tabular}{|c||c||c|} \hline
\multirow{2}{*}{System} & \multicolumn{2}{c|}{Attribute}	\\ \cline{2-3}
					& Emph 		& Quality \\ \hline\hline
Base (NoEmph)	& 2.20 (1.3)	& 3.87 (0.9)	\\ \hline
Base (Sup)		& 3.71 (1.2)	& 3.97 (0.9)	\\ \hline
PC-Unsup			& 3.58 (1.1)	& 3.97 (0.9)	\\ \hline
PC-Hybrid			& 4.02 (1.0)	& 4.08 (0.8)	\\ \hline
\end{tabular}
\end{table}

\section{Discussion and Conclusions}
\label{sec:discuss}

From these evaluations, we can make the following remarks for both speakers. All controllable systems achieved
a much higher degree of emphasis than {\em Base (NoEmph)} (which, as expected, attained low scores in
terms of emphasis realizability), and this was achieved at no expense of overall quality since the remaining systems
are statistically better or the same. We hypothesize this improvement in quality is due to the fact that conditioning 
on additional prosodic attributes of the outputs steers the model toward more natural (and stable) points during training.
We observe differences between the approaches, however,
comparing the {\em matched} vs. {\em transplant} conditions: when labeled data is available for a target speaker, our experiments
suggest that the fully-supervised approach offers the best operating point in terms of both quality and emphasis (Table~\ref{table:lt-jjw}). 
This approach, however, does not generalize as well as the hybrid approach does to a new target speaker lacking labeled data  
(Table~\ref{table:lt-sab}). For the latter, combining  
supervision with the prosodic-conditioning framework supplements the performance for both attributes when training a multi-speaker
model to enable the transfer of knowledge. Lastly, we see that even lacking any labeled data,
the framework is able to provide a good point of quality and emphasis control by means of boosting the predictions of the fully
unsupervised model. This is facilitated by our use of a set of controls 
that are readily interpretable and can be perceptually linked to the task at hand.
Though the results are very encouraging, some difficult test cases remain. For instance, we have observed in informal listening 
the challenge posed by some function words, particularly clitics or words containing only
unstressed vowels in broad-focus realizations.

We have introduced and validated a framework that allows for a finer degree of control
over  lexical prosody to guide the realization of narrow focus in S2S synthesis. This framework encompasses a set of user-driven controls that
meet the criteria that we highlighted and advocated for in Sec.~\ref{sec:intro} of the paper: they consist of a low-dimensional representation of prosody,
they are intuitive in the sense that changes to the controls map to identifiable perceptual effects in the output, and they offer a mechanism that
disentangles different components of prosody (duration and pitch) that can be tuned separately.
The approach requires only a moderate amount of knowledge
external to the framework in the form of coarse word-level alignments, and we have shown that it can accommodate 
various degrees of supervision
depending on available resources, with different variants bringing in different strengths depending on the operating conditions (e.g.,
synthesizing from a speaker with labeled supervised data vs. transplanting to a novel speaker that lacks such resources).

We should note that this framework can also be extended to include other levels of the prosodic
hierarchy to explore expressive effects beyond localized narrow focus. For instance,
incorporating the intonational phrase into the analysis might provide a way to better model the pitch reset associated
with parentheticals. Addressing the shortcomings already mentioned and incorporating these extensions remain the subject of ongoing and future work.

% Below is an example of how to insert images. Delete the ``\vspace'' line,
% uncomment the preceding line ``\centerline...'' and replace ``imageX.ps''
% with a suitable PostScript file name.
% -------------------------------------------------------------------------
%\begin{figure}[htb]
%
%\begin{minipage}[b]{1.0\linewidth}
%  \centering
%  \centerline{\includegraphics[width=8.5cm]{image1}}
%%  \vspace{2.0cm}
 % \centerline{(a) Result 1}\medskip
%\end{minipage}
%
%\caption{Example of placing a figure with experimental results.}
%\label{fig:res}
%
%\end{figure}

%\vfill
%\pagebreak


\bibliographystyle{IEEEbib}
\bibliography{emph_slt2021_arXiv}
%\bibliography{C://Papers//Bib//nn,C://Papers//Bib//prosody,C://Papers//Bib/prosody_AMR,C://Papers//Bib//learning,C://Papers//Bib//tts,C://Papers//Bib//ttse2e,C://Papers//Bib//voicequal,C://Papers//Bib//optimization,C://Papers//Bib//url}

\end{document}